\documentclass[11pt]{article}
\usepackage{epsfig}
\usepackage{graphicx,epsfig}
\long\def\AUTHORS#1{ #1\\[3mm]}
\long\def\AFFILIATION#1#2{$^{#1}\,$ #2\\}
\begin{document}
\vspace{12pt}
\def\Title#1{\begin{center} {\Large {\bf #1} } \end{center}}

\Title{Present status of the search for the K$^{0}_{L} \rightarrow \pi^{0}\nu\bar{\nu}$ decay with the  KOTO detector at J-PARC$^2$\\
}

\bigskip\bigskip

\AUTHORS{\underline{B. Beckford}$^1$ on behalf of the KOTO collaboration }
{\small \it 
\AFFILIATION {1}{Department of Physics, University of Michigan, Ann Arbor, MI 48109 USA}}
\AFFILIATION {2}{Talk presented at the APS Division of Particles and Fields Meeting (DPF 2017), July 31-August 4, 2017, Fermilab. C170731}

\centerline{Contact email: {\it bobeck@umich.edu}}

\begin{center}
\section*{Abstract}
\end{center}
We have performed a search for the K$^{0}_{L} \rightarrow \pi^{0}\nu\bar{\nu}$ decay with the KOTO detector at J-PARC. The KOTO detector was designed to observe the decay and measure its branching ratio (BR). Focusing on this \emph{golden} decay in quark flavor physics provides an ideal candidate to probe for physics beyond the standard model (BSM). The established experimental upper limit of the branching ratio was set by the KEK E391a collaboration at 2.6 x 10$^{-8}$. This is still well above the standard model value of the branching ratio, which is predicted to be 2.43 x 10$^{-11}$ with minor uncertainties.

The important signal is a pair of photons from the $\pi^{0}$ decay and no coincident signals from veto counters. This along with a large discernible transverse momentum provides us with unique signature. KOTO uses a Cesium Iodide (CsI) electromagnetic calorimeter as the main photon detector and hermetic veto counters to ensure that there are no other detected particles. This proceeding summarizes a presentation that discussed achievements, improvements to the detector, detailed the current analysis status for 2015 data, and future prospects.\\ 

\section{Introduction and motivation}
Investigation into CP violation along with deepening our comprehension of neutrino mixing remains a major area of study in elementary particle physics and continues to gain attention experimentally and theoretically. The $\emph{golden}$ decay of K$^{0}_{L} \rightarrow \pi^{0}\nu\bar{\nu}$ presents one of the cleanest avenues for understanding the combined violation of charge and parity as the decay rate proceeds almost entirely from a direct CP violating amplitude. The transition from a strange to a down quark, (s$\rightarrow$ d), $\Delta$ S = 1, is prohibited at the tree level and takes places through higher order penguin or box diagrams which is highly suppressed in standard model calculations~\cite{Littenberg}. 
Using the Wolfenstein parameterization of the Cabibbo-Kobayashi-Maskawa (CKM) matrix permits an expression of the matrix elements with respect to the unitary triangle. A consequence of unitarity is a decay rate proportional to the square of $\eta$ and the height of the unitary triangle. 

A model-independent branching ratio limit of K$^{0}_{L} \rightarrow \pi^{0}\nu\bar{\nu}$ was set from isospin symmetry arguments placed on the BR(K$^{+} \rightarrow \pi^{+}\nu\bar{\nu}$) of (17 $\pm$ 11 $\pm$ 10) x 10$^{-11}$. The BR(K$^{+} \rightarrow \pi^{+}\nu\bar{\nu}$) was reported by the BNL E949 experiment~\cite{Artamonov}. This Grossman-Nir bound has constrained the branching fraction to BR(K$^{0}_{L} \rightarrow \pi^{0}\nu\bar{\nu})  < 1.46$ x $10^{-9}$~\cite{Grossman}. Currently the SM decay rate prediction is (2.43 $\pm$ 0.309 $\pm$ 0.06) x 10$^{-11}$. An experimental upper limit of 2.6 x 10$^{-8}$ at the 90\% confidence level (C.L.) was placed by the KEK E391a experiment~\cite{Brod,Buras,Ahn}. 

In spite of the experimental challenges associated with a measurement involving all neutral particles in the final state, this decay continues to provide one the best methods for investigating CP violation in the quark sector. Observation of the decay rate within the uncertainties of prediction would further support the SM description conversely, a measured rate above the SM prediction would serve as an indicator of new physics beyond the standard model (BSM)~\cite{Littenberg}. Growing interest in rare kaon decays is evident not only in the efforts of the KOTO experiment but can be seen through additional searches being performed at CERN. The NA62 experiment at CERN is aimed at measuring charged kaon decays (K$_{L}^{+} \rightarrow \pi^{+}\nu \bar{\nu}$)~\cite{CERNKPLUS}. 
Both  KOTO and NA62 results will be pivotal to advancing our understanding of the underlying CP-violating process.

\section{KOTO Experiment}
\subsection{KOTO detector}
\label{sec:Detector}
  \begin{figure}[htb]
	\begin{center}
	\epsfig{file=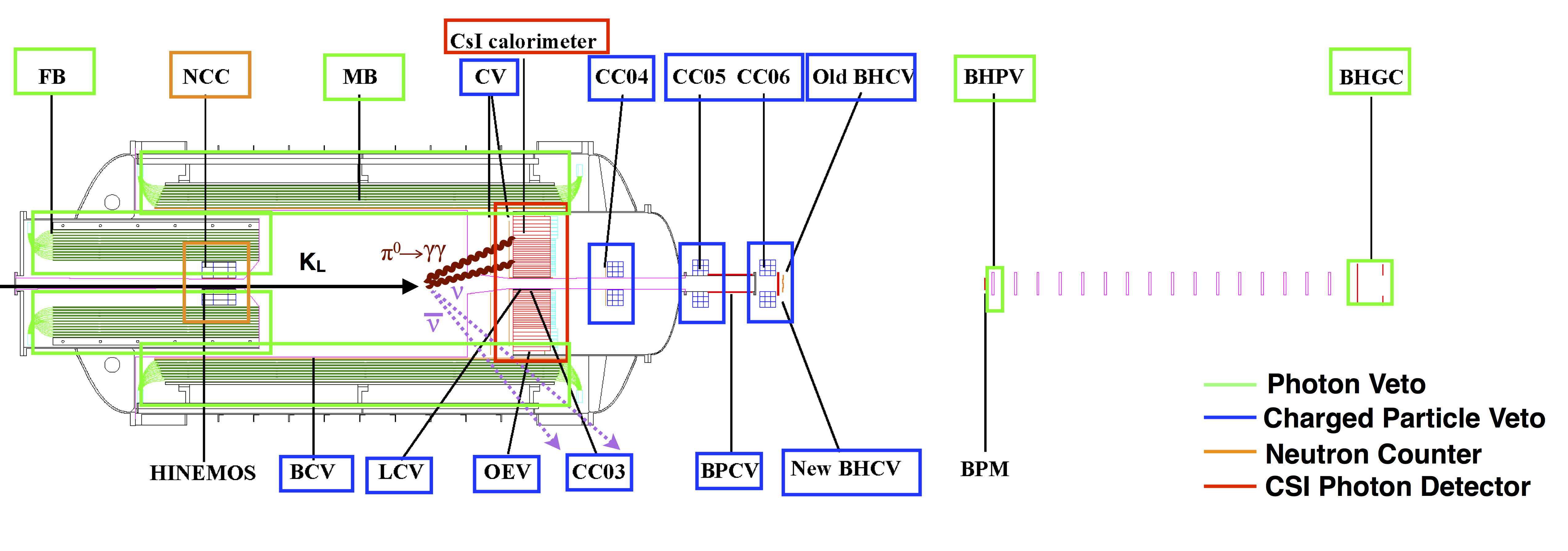,width=1.0\columnwidth}
			\caption{(Color online) Cross-sectional side view of the KOTO detector. The beam enters from the left hand side and goes down along the z axis.}
  	\vspace{-.5cm}
	\label{fig:koto_schematic}
	\end{center}
\end{figure}
The KOTO experiment is performed at the J-PARC Hadron Experimental Facility (HEF) and aims to observe the K$^{0}_{L} \rightarrow \pi^{0}\nu\bar{\nu}$ decay via the direct measurement of $\pi^{0}\rightarrow \gamma\gamma$. Kaons were generated from a beam of 30-GeV protons that are slowly extracted from the Main Ring (MR) with stable beam intensity. The proton beam strikes on a 66-mm-long gold target at HEF and produces the kaons that are directed at a 16 degree angle from the primary proton beam to the KOTO detector. The extraction at this angle reduces the energies of the neutrons that are part of the neutral beam thus reducing background for our measurement. The neutral beam is collimated to achieve a so-called pencil beam with a solid angle of 7.8 $\mu$sr and a size of 5 x 5 cm$^2$  downstream of the target. This size constraint was used as an additional restriction in determining  the decay position in the transverse direction to the beam. We use the term \emph{halo neutrons} to describe neutrons located outside the nominal beam solid angle.  

Figure~\ref{fig:koto_schematic} shows a cross-sectional side view of the KOTO detector. The detector is a two sub system design. The electromagnetic calorimeter is composed of 2716 Cesium Iodide (CsI) crystals is used to measure the positions and energies of two photons from $\pi^{0}$ decay. All other detectors are used as a charged particle or a photon veto system. The K$^{0}_{L}$ decay volume along the direction of the beam has an expectation length of 3 meters.  


\subsection{Experimental method}
A direct measurement of K$^{0}_{L} \rightarrow \pi^{0}\nu\bar{\nu}$ is unfeasible, thus, the decay signature is a single $\pi^{0}$ from a K$_L$ decay with the neutrinos going undetected. The veto system described in Section~\ref{sec:Detector} provides a hermetic veto to detect charged particles or photons from kaon decays andother sources. Due to the undetected neutrinos we expect the two photons from the $\pi^{0}$ decay to be reconstructed with a large transverse momentum. Experimentally this decay mode poses challenges in being observed. A primary obstacle lies in the rejection of background from other K$^{0}_{L}$ decays, such as  K$_{L} \rightarrow \pi^{0}\pi^{0}$, K$_{L} \rightarrow \pi^{+}\pi^{-}\pi^{0}$, and \emph{halo neutron} induced shower events in the CsI crystals, while maintaining sensitivity to the K$^{0}_{L} \rightarrow \pi^{0}\nu\bar{\nu}$ process.

\section{Results}
         \begin{figure}[ht]
         \begin{center}
   	\epsfig{file=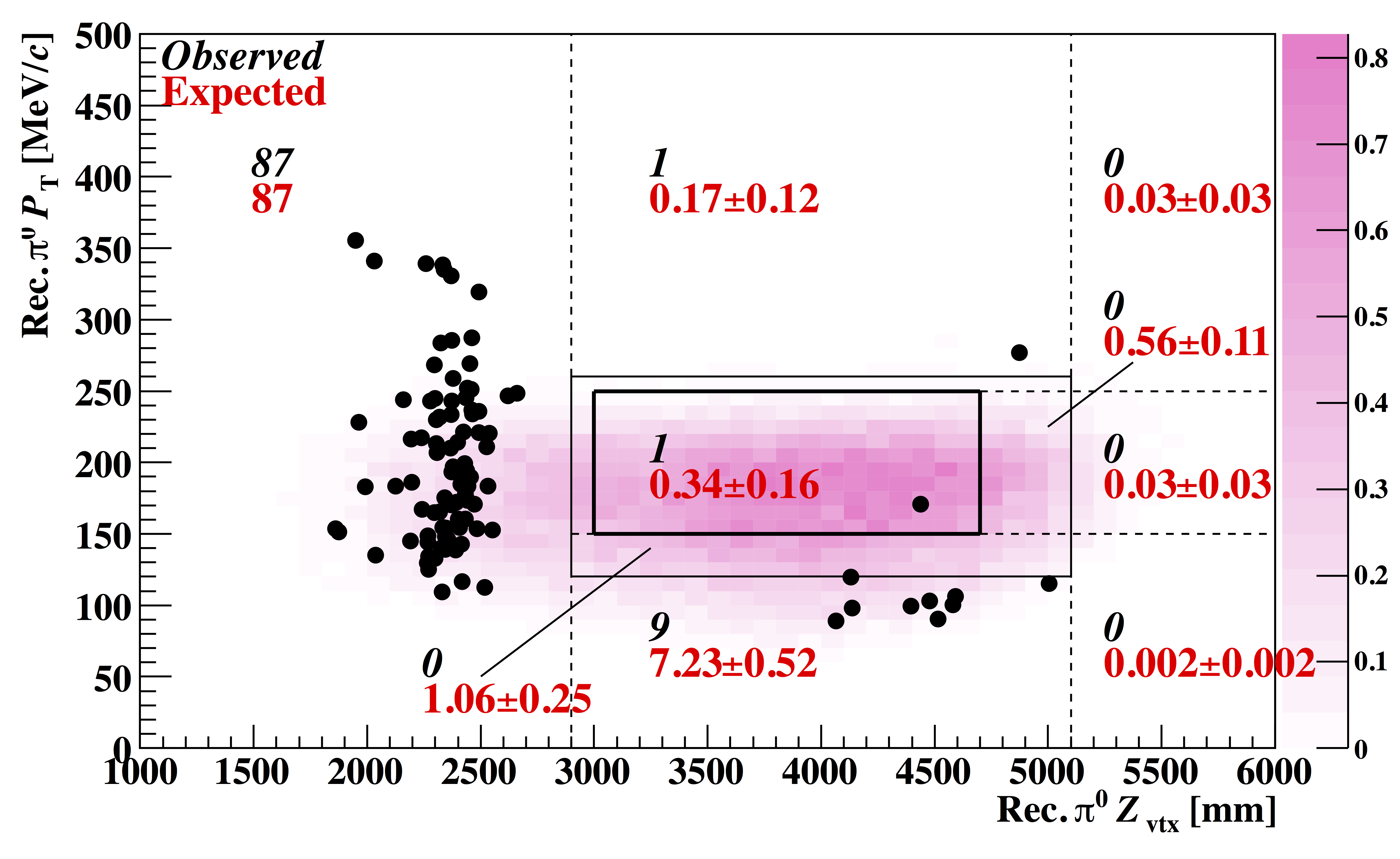,width=0.650\columnwidth}
            \caption{(Color online) Reconstructed $\pi^{0}$ transverse momentum versus decay vertex for events after all analysis stipulations from the 2013 data set. The expected signal region is depicted by the pink shaded area determined from Monte Carlo. The blinded and signal region are shown as the thin solid and bold rectangular areas. Event expectations for data and Monte Carlo in different regions are shown as the black and red numbers, respectively~\cite{Maeda}.}
          \label{fig:2013_results}
                  \end{center}
    \end{figure}
   
The collaboration performed a search for $K_L\rightarrow \pi^{0}\nu\bar{\nu}$ with the initial physics data taken in 2013 (run49). The experimental period was shortened as a result of an accelerator accident limiting the data gathered to 100 hours of Protons on Target (POT). In Fig.~\ref{fig:POT}, the accumulated protons on target (POT) for the experimental runs and the corresponding beam power are given. From our blind analysis there was one event observed with the expectation of 0.34 $\pm$ 0.16 background events. Figure~\ref{fig:2013_results} shows the reconstructed $\pi^{0}$ transverse momentum (P$_t$) versus reconstructed decay vertex (right) for candidate events after the application of all offline analysis cuts for the inaugural physics run (run49). The smaller rectangular area is the K$^{0}_{L} \rightarrow \pi^{0}\nu\bar{\nu}$ signal box. Expectations of events for data and Monte-Carlo in different regions are shown as the black and red numbers, respectively, in the figure. A summary of estimated background events in the signal region are listed in Table~\ref{background_table_run49}. Based on the observation of one event in the signal region, the collaboration set an upper limit of 5.1 x 10$^{-8}$ for the $K_L\rightarrow \pi^{0}\nu\bar{\nu}$ branching fraction at the 90\% confidence level (C.L.)~\cite{Maeda}.

 
    \begin{table}[h!]
\caption{ Summary of estimated background events in the signal region }
\begin{center}
\begin{tabular}{|c|c|}
  \hline
  Source of background & Estimated events \\
  \hline
	K$^{0}_{L}$ decay events	& 0.10 $\pm$ 0.04 \\
	Halo neutron events on CsI & 0.18 $\pm$ 0.15 \\
	Halo neutrons events on NCC	& 0.06 $\pm$ 0.06 \\
  	\hline
   	\hline
   	Total	& 0.34 $\pm$ 0.16		\\
	\hline
	\hline
\end{tabular}
\end{center}
\label{background_table_run49}
\end{table}

\subsection{Detector upgrades }
The main background contributions in the signal region came from halo neutron events, where the neutron struck the CsI and the resulting showers mimicked two photon events, K$_{L} \rightarrow \pi^{0} \pi^{0}$, and K$_{L} \rightarrow \pi^{+}\pi^{-}\pi^{0}$ events. In the latter cases, charged particles and extra photons went undetected in the main barrel or escaped through the beam hole located at the center of the CsI resulting in only two photon events being detected. To reduce these events we have included the following detectors: the Inner Barrel (IB), the Beam Hole Charge Veto (BHCV), the Beam Hole Photon Veto (BHPV), and the Beam Hole Guard Counter (BHGC). The Inner Barrel was added primarily to increase photon detection in the barrel region. The current beam power of 45-50 kW is expected to be increased to 80 kW. Also, we replace the current target. Additional planned detector improvements involve the installation of Multi Pixel Photon Counters (MPPC). The timing information from the combination of MPPCs and PMT affixed to the CsI crystals will be used to distinguish between neutron and photon events. The data acquisition system (DAQ) will also be upgraded to accommodate the increased beam power. Details concerning DAQ upgrades can be found in reference~\cite{Su}. 

 	\vspace{-.3cm}

\section{Summary and outlook }
   \begin{figure}[ht]
     \begin{center}
    	 \epsfig{file=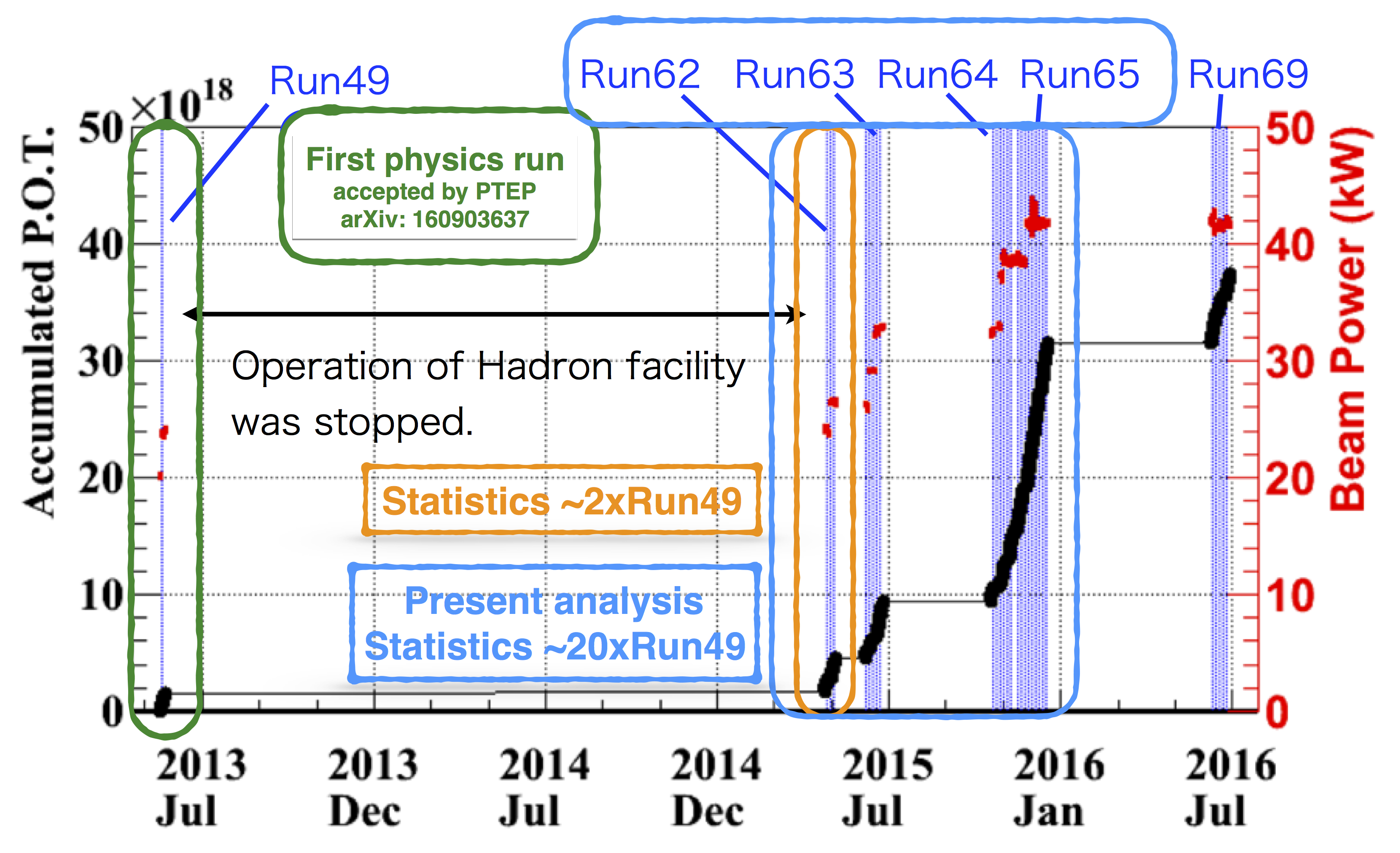,width=0.770\columnwidth}
            \caption{(Color online) Accumulated Protons on Target (POT) since the launch of the KOTO experiment is shown on the left axis. The right axis shows the associated beam power for each run.}
          \label{fig:POT}
        \hspace{0.05\linewidth}
        \end{center}
            \end{figure}
            
\begin{figure}[h!]
	\begin{center}
		\epsfig{file=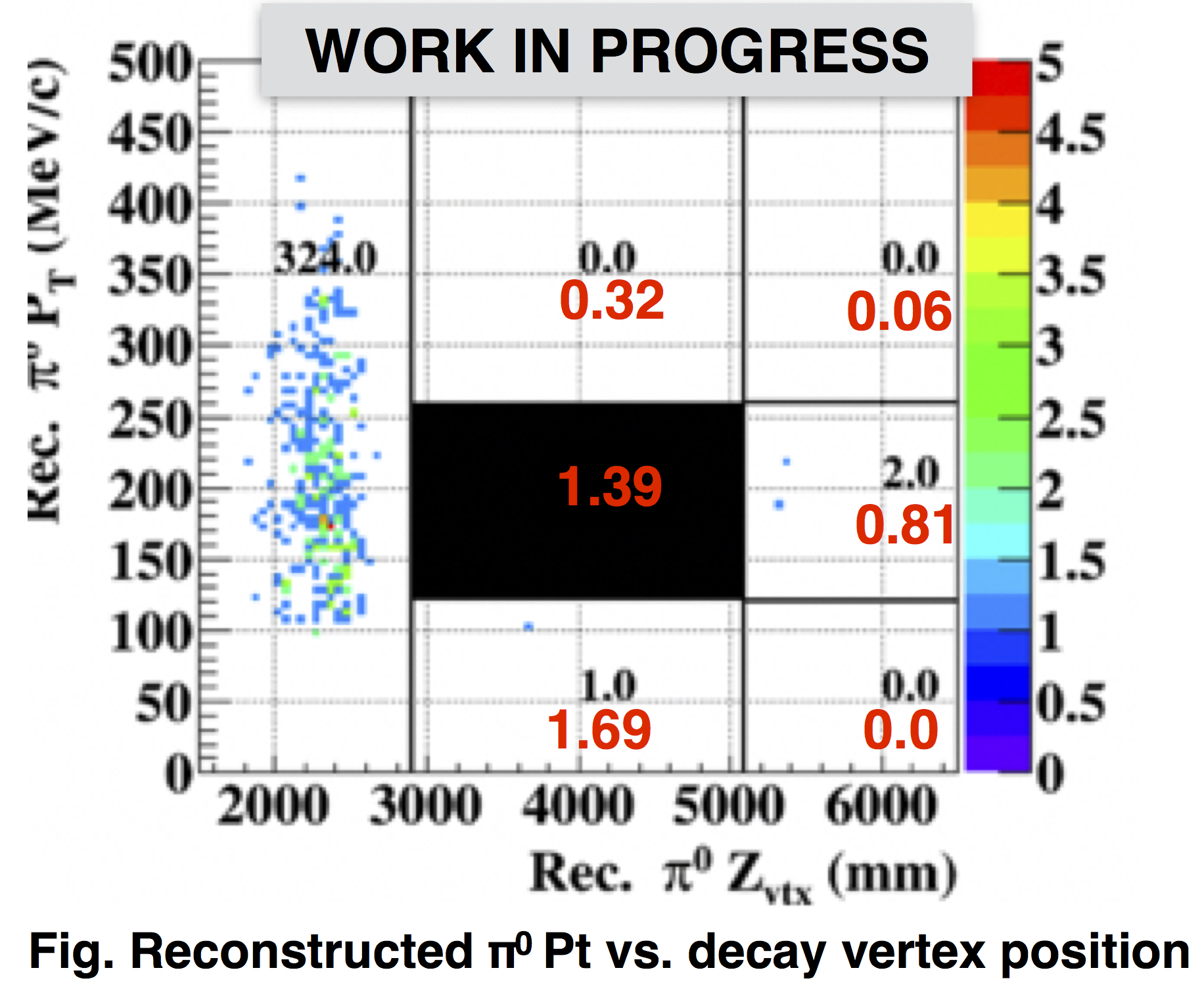,width=0.650\columnwidth}
			\caption{(Color online) Reconstructed $\pi^0$ transverse momentum versus reconstructed decay vertex for candidate events from the 2015 data set with kinematic and veto cuts applied. The shaded rectangular area is the K$^{0}_{L} \rightarrow \pi^{0}\nu\bar{\nu}$ blinded signal box. Observed events for data and expectations from Monte-Carlo in different regions are shown as the black and red numbers, respectively, in the figure.}
  	\vspace{-.2cm}
	\label{fig:2015_results}
	\end{center}
\end{figure}

The data taken during 2015 (Runs 62$-$65), at beam powers of 24-42 kW as shown in Figure.~\ref{fig:POT}, when finalized is expected to approach sensitivity of the Grossman-Nir Bound. Figure~\ref{fig:2015_results} presents the reconstructed $\pi^0$ P$_t$ versus decay vertex for this data set. The shaded rectangular area is the K$^{0}_{L} \rightarrow \pi^{0}\nu\bar{\nu}$ blinded signal box. Events expectations of background events for data and Monte Carlo are shown as the black and red numbers, respectively, in the figure. A listing of the estimated background contributions and sources are provided in Table~\ref{table:background_table_run62}. Similar to the results of the 2013 run, the major background contribution is associated with neutron induced events. 

\begin{table}[h!]
\caption{Preliminary summary of estimated background events in the signal region}
\begin{center}
\begin{tabular}{|l|c|}
  \hline
  Source of background & estimated events \\
  \hline
	K$^{0}_{L}\rightarrow \pi^{0}\pi^{0}$			& 0.07 $\pm$ 0.07 \\
	K$^{0}_{L}\rightarrow \pi^{0}\pi^{+}\pi^{-}$	& 0.23 $\pm$ 0.06 \\
	Neutrons events on NCC (upstream)			& 0.13 $\pm$ 0.03 \\
	Halo neutron events hitting CsI				& 0.34 $\pm$ 0.11 \\
	 CV $\pi^{0}$ 								& 0.14 $\pm$ 0.03 \\
	 CV $\eta$									& 0.48 $\pm$ 0.05 \\

  	\hline
   	\hline
   	Total	& 1.39 $\pm$ 0.15	\\
	\hline
	\hline
\end{tabular}
\end{center}
\label{table:background_table_run62}
\end{table}

We have successfully performed a search for K$^{0}_{L} \rightarrow \pi^{0}\nu\bar{\nu}$ with the KOTO detector. Our detailed analysis of data obtained  in 2013 (run 49) yielded one observed event in the blinded signal region with an expectation of 0.34 $\pm$ 0.16 background events. Consequently, we placed an upper limit for the branching fraction of 5.1 x 10$^{-8}$ at a 90\% confidence level (C.L.). 
Due to this foundational achievement we are continuing the search by making improvements that have included upgrades to the detector and the data acquisition system. Since restarting the search in 2015, we have accumulated roughly 20 times more data than the initial run. The KOTO collaboration will continue analyzing the K$^{0}_{L}$ events from the normalization modes (K$^{0}_{L} \rightarrow \pi^{0}\pi^{0}\pi^{0}$, K$^{0}_{L} \rightarrow \pi^{0}\pi^{0}$, and K$^{0}_{L} \rightarrow \gamma\gamma$) in order to determine the K$^{0}_{L}$ flux and normalize the single event sensitivity, and improve cut selections in order to further reduce background contributions.


\section{Acknowledgements}
This work supported by the Department of Energy under grant DE-SC0007859.

\def\Discussion{
\setlength{\parskip}{0.3cm}\setlength{\parindent}{0.0cm}
     \bigskip\bigskip      {\Large {\bf Discussion}} \bigskip}
\def\speaker#1{{\bf #1:}\ }
\def\endDiscussion{}





 

\begin{thebibliography}{99}

\bibitem{Artamonov} A. V. Artamonov et al., Phys. Rev. Lett, vol. 101, p. 191802, 2008.
\bibitem{Grossman} Y. Grossman et al., Phys. Lett. $\bf{B398}$ 163 (1997)
\bibitem{Brod} J. Brod et al., Phys. Rev. $\bf{D83}$, 034030 (2011)
\bibitem{Buras} A. J. Buras, D. Buttazzo, R. Knegjens, J. High Energy Phys. 1511, 166 (2015).
\bibitem{Ahn} J. Ahn et al., Phys. Rev. $\bf{D81}$, 072004 (2010)
\bibitem{Littenberg} L. Littenberg, Phys. Rev. $\bf{D81}$, 3322 (1989) 
\bibitem{CERNKPLUS} CERN-SPSC-2005-013 SPSC-P-326
\bibitem{Maeda} J. Ahn et al., (PTEP) $\bf{021C01}$; (2017)
\bibitem{Su} S. Su. et al., IEEE Transaction of Nuclear Science Vol 64, No. 6.



\end{thebibliography}
\end{document}